\documentclass{aa}  

\usepackage{graphicx}
\usepackage{txfonts}
\usepackage{natbib}
\bibpunct{(}{)}{;}{a}{}{,}
\usepackage{hyperref}
\hypersetup{
       hyperfootnotes=true,			
	   bookmarks=true,			
	   colorlinks=true,
	   linkcolor=black,
       linktoc=page,
	   anchorcolor=black,
	   citecolor=blue,
	   urlcolor=black,
}
\usepackage{amsmath}
\begin{document}

   \title{Mass assembly and AGN activity at $z\gtrsim1.5$ in the dense environment of XDCPJ0044.0-2033}


   \author{M. Lepore
          \inst{1,2}
          \and
          A. Bongiorno \inst{2}
          \and
          P. Tozzi\inst{1}
          \and
          A. Travascio \inst{3}
          \and
          L. Zappacosta \inst{2}
          \and 
          E. Merlin\inst{2}
        \and 
        R. Fassbender\inst{4}
          }

   \institute{INAF-Osservatorio Astronomico di Arcetri, Largo Enrico Fermi 5, 50125, Florence, Italy.\\
              \email{marika.lepore@inaf.it}
         \and
             INAF-Osservatorio Astronomico di Roma, Via di Frascati 33, 00078, Monteporzio Catone, Rome, Italy.
        \and
             Università degli Studi di Milano Bicocca, Piazza dell'Ateneo Nuovo 1, 20126, Milan, Italy.
         \and 
              Max Planck Institut f\"ur extraterrestrische Physik, Giessenbachstrasse 1, 85748 Garching, Germany.
             }


 
  \abstract
  {XDCPJ0044.0-2033 is the most massive galaxy cluster known at $z>1.5$ 
  and its core shows a high density of galaxies which are experiencing 
  mergers and hosting nuclear activity.}
   {We present a multi-wavelength study of a region of 24 kpc $\times$ 24 kpc 
   located $\sim 157$ kpc from the center of the cluster, for which we have 
   photometric and spectroscopic observations.
   Our main goal is to investigate the 
   environmental effects acting on the galaxies inhabiting this
   high density region.}
   {We performed sources identification and photometric analysis 
   on high-resolution Hubble Space Telescope (HST) images 
   in F105W-, F140W- and F160W-band and spectroscopic analysis 
   of the Near-Infrared (NIR) KMOS data in H and YJ bands. 
   In addition, we analyzed the deep \emph{Chandra} ACIS-S X-ray exposure.}
   {We find that the analyzed region hosts at least 9 different sources, 
   6 of them confirmed to be cluster members within a narrow redshift range 
   $1.5728<z< 1.5762$, and therefore is denser than the very central, more massive 
   region of the cluster previously analyzed by \cite{Travascio2020}. 
   These sources form two different complexes (\texttt{Complex M} 
   and \texttt{Complex N}) at a projected distance of $\sim 13$ kpc, 
   which are undergoing merging on an estimated timescale of 10-30 Myr.  
   One of the sources shows the presence of a broad H$\alpha$ emission line and 
   is classified as Type-1 Active Galactic Nucleus (AGN). 
   This AGN is associated to an X-ray point-like source, 
   whose emission appears moderately obscured 
   ( with intrinsic absorption N$_{H}\sim10^{22} cm^{-2}$) 
   and hosts a relatively massive black hole (BH) with mass $M_{BH}\sim 10^{7} M_{\odot}$, 
   which is accreting  with an Eddington ratio of $\sim 0.2$.}
   {We conclude that the analyzed region is consistent with being the formation site 
   of a secondary Brightest Cluster Galaxy (BCG). These findings, together with 
   an in-depth analysis of the X-ray morphology of the cluster, suggest a merging scenario
   for the entire cluster, with two massive halos both harboring two rapidly evolving 
   BCGs on the verge of being assembled. Our results are also consistent 
   with the scenario in which the  AGN phase in member galaxies is triggered 
   by gas-rich mergers, playing a relevant role in the formation of the 
   red sequence of elliptical galaxies observed in the center of local galaxy clusters.}


   \maketitle

\section{Introduction}
\label{sec:Intro}

The standard $\Lambda$CDM cosmological model predicts a hierarchical scenario 
for the formation and evolution of large scale structures in an expanding Universe 
\citep{Bond1991}. In this scenario, the first structures in the Universe are halos 
of Dark Matter (DM) that, through merging, form increasingly massive halos. 
The baryonic matter then falls into the potential wells of these DM halos and, 
through gas accretion and/or merging, forms stars and galaxies \citep{WhiteReese1978}.\\ 
Eventually, the environment in which a galaxy is located influences its evolution. 
In particular, this evolution depends on the presence of cold molecular gas and, 
in dense environments, processes as galaxy-galaxy mergers 
\citep{AarsethFall1980,ParkHwang2009}, harassement \citep{Moore1996}, 
strangulation \citep{BaloghNavarroMorris2000, vandenBosch2008} and ram-pressure stripping 
\citep{GunnGott1972,McCarthy2008} can occur and affect the transportation of the cold 
gas thus accelerating the evolution of a galaxy. \\
Galaxy clusters are the ideal laboratories for studying the effects of dense environments 
on the evolution of galaxies and their properties, 
especially in their early phases when gas-rich, rapidly forming galaxies 
strongly interact with each other. In this framework, the study of galaxy clusters at 
different redshift is key to understand the role of the environment 
in the evolution/transition from star forming (SFGs) to red and passive galaxies 
\citep{Alberts2016}. 

On one hand, in the local Universe and up to $z\sim 1.4$, 
the dense cores of galaxy clusters are preferentially populated by massive early-type galaxies that form a tight sequence in the color-magnitude diagram (CMD), the so called Red Sequence \citep{Bell2004}. Among them, the BCG, typically a giant elliptical, stands out for its high luminosity ($M_{V}\sim-23$) and stellar mass ($M\sim10^{12}M_{\odot}$). 
While at $z<1.4$ the SFGs are preferentially located in the cluster outskirts, where the local galaxy density is low, a reversal of the star formation (SF)-density relation starts to appear at $z\gtrsim 1.4$.  Therefore, massive, high-redshift galaxy clusters host more SFGs in their core \citep{Brodwin2013}, forming the so called Blue Sequence \citep{Bell2004}, and may not show the presence of a BCG.  This inversion suggests that the galaxy cluster populations undergo one or more processes able to affect the SF activity in a relatively short time. 

Also the AGN activity in clusters shows a similar trend compared to the SFGs. 
As an example, \cite{Alberts2016}, studying SF and AGN activity in 11 galaxy clusters at $1<z<1.75$, suggested a co-evolution between SF and AGN driven by merger activity. The current framework, in agreement with semi analytic models and simulations, predicts a scenario in which galaxy major mergers induce starburst and also fuel BH accretion, triggering an AGN phase \citep{Hopkins2006}.
Later, the feedback phase of AGNs, in form of winds and outflows, lead to SF quenching and to the formation of passive elliptical galaxies, as observed in clusters in the local Universe \citep{Narayanan2010}. However, it is not clear yet how these processes happen and how galaxy properties can be influenced by the presence of nuclear activity and vice versa. In this context, $z\sim 12$ is a crucial epoch to observe young galaxy clusters, where we do expect a high rate of mergers among galaxies and a higher rate of 
nuclear and SF activity. 

Over the years, several large area surveys have been designed to reveal high redshift, massive ($M\gtrsim10^{14}M_{\odot}$) galaxy clusters
using optical/infrared/X-ray observations. Among them: SpARCS \citep{Wilson2006} is based on the detection of distant clusters by means of their red galaxy population; SPT \citep{Williamson2011} and ACT \citep{Menanteau2010} are based on the Sunyaev-Zel'dovich effect (SZ); and the XMM-Newton Distant Cluster Project \citep[XDCP,][]{Fassbender2011b} is based on the study of diffuse X-ray emission from the Intracluster Medium (ICM). \\
With these surveys it was possible to state that the SF-density relation gradually changes starting from $z>1$, (e.g. \citealp{Hilton2010,Fassbender2011b,Tadaki2012,Fassbender2014,Santos2014,Santos2015}). Also, a decline in SF in clusters over cosmic time \citep{Alberts2014}, paralleled by a decrease in black hole activity has been unambiguously observed and
quantified. The fraction of AGNs falls by two orders of magnitude in clusters from $z \sim 1.5$ to $z \sim 0$ (\citealp{Galametz2009,Martini2013}), with AGNs preferentially residing in the infall regions of clusters \citep{Pimbblet2013}. 
Quantifying, at the same time, the effects of cosmic evolution and the 
environmental effects is, therefore, a key aspect of the evolution of massive galaxies.

\begin{figure*}
 \centering
            {\includegraphics[scale=0.9]{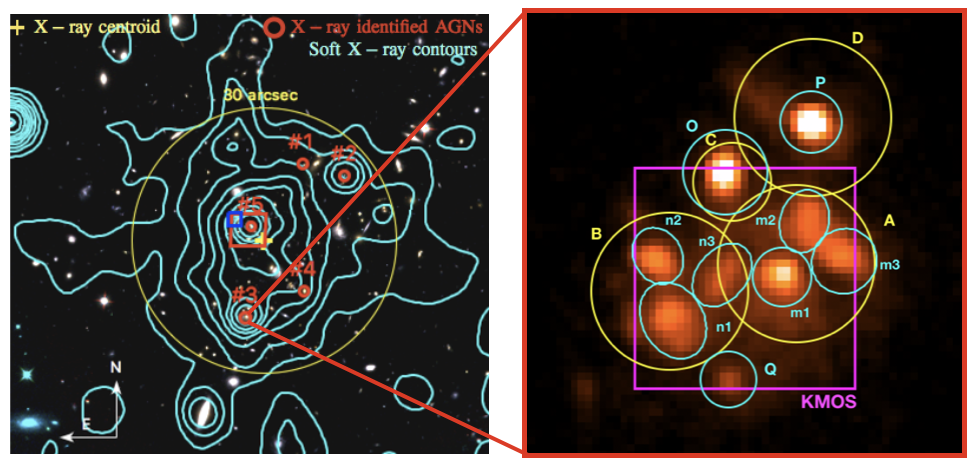}}
      \caption{\emph{Left Panel}, adapted from \citealp{Travascio2020}: HST RGB (F105W+F104W+F160W) image of \texttt{XDCP0044} with overlaid the soft ([0.5-2] keV) bands X-ray contours. The yellow circle is centered on the
      X-ray extended emission centroid (yellow cross, RA=00:44:05.2 and DEC=-20:33:59.8) and has a radius of
      30 arcsec, corresponding to $\sim 250$ kpc. The red circles mark the 5 X-ray point-like sources detected by \cite{Tozzi2014}.  \emph{Right Panel}: HST image in F105W band. The yellow circles mark the sources detected in ground based HAWK-I J and Ks bands by \cite{Fassbender2014} while cyan circles mark the sources detected in HST. The magenta square is the KMOS field of view.}
         \label{fig:clusterreg3}
   \end{figure*}

In this paper we focus on the high-z galaxy cluster XDCP J0044.0-2033 (or \emph{"Gioiello"} cluster, hereafter \texttt{XDCP0044}; \citealp{Fassbender2011b}, \citealp{Santos2011}), the most massive 
cluster known at $z>1.5$ (see Figure \ref{fig:clusterreg3}, \emph{left panel}). 
It has been found thanks to its extended X-ray emission, discovered in the XDCP survey \citep{Fassbender2011b}. Deep \emph{Chandra} observations confirmed strong diffuse emission (with centroid coordinates $RA=00:44:05.2$ and $DEC=-20:33:59.7$) typical of virialized clusters \citep{Tozzi2014}.
\texttt{XDCP0044} is thus in a quite advanced dynamical state and is an ideal laboratory to study the mass assembly at high-z, the interplay between galaxies, nuclear activity, and the intergalactic gas in dense environment. 

Concerning the galaxy population, \texttt{XDCP0044} shows a strong inversion of the SF-density relation. Indeed, \cite{Santos2014} found that the total SF rate within the projected core area ($r<250$ kpc) is SFR=1875 $\pm$ 158 $M_{\odot}/yr$ (considering only the spectroscopic members), four times higher compared to the SFR in the cluster outskirts. 
In addition, X-ray point-like emission has been detected from at least 2 member galaxies within $\sim$ 250 kpc from the cluster center (labeled as 3 and 5 in Figure \ref{fig:clusterreg3}, \emph{left panel}). 
Two additional unresolved X-ray sources may turn out to be additional cluster members, but
they do not have redshift measurements yet.
Furthermore, radio JVLA data show the presence of at least four sources emitting in the radio band and an extended radio emission associated to \texttt{Complex A} \citep[see][]{Travascio2020}. 
In particular, the core region ($\sim$ 70 kpc $\times$ 70 kpc) of \texttt{XDCP0044}, studied by \cite{Travascio2020}, is the densest among those observed at $z>1.5$ and shows the presence of two luminous, obscured and highly accreting AGNs and an optically obscured Type-2 AGN, with signatures of ongoing merging. Multiple AGN activity and high SF in the core of a high-z cluster suggest that these processes have a key role in shaping the nascent BCG observed at the center of local clusters. 

On the basis of these results, we decided to investigate star formation and nuclear activity in another region located $\sim 157$ kpc from the core of \texttt{XDCP0044}, that appears to be extremely 
dense and active, as shown by the presence of an X-ray unresolved source at its center, the intense star formation at the level of SFR=825$\pm$120 $M_{\odot}/yr$ \citep[source ID 95 in][]{Santos2014}, and
the presence of a compact radio source with luminosity $L_{1.5 GHz}=2.6 \pm 0.3 \times 10^{24} W/Hz$ \citep[source r3 in][]{Travascio2020}. 
To perform our study, we use high-resolution HST 
photometric data combined with K-band multi-object spectrometer at ESO VLT
(KMOS IFU) observations and X-ray \emph{Chandra} data.
Figure \ref{fig:clusterreg3} shows the HST RGB (F105W+F140W+F160W) image of \texttt{XDCP0044} with overlaid the soft ([0.5-2.0] keV) band X-ray contours of the extended emission and the 5 X-ray point-like sources detected by \cite{Tozzi2014}. 

The paper is organized as follows. In Section \ref{sec:obs and data} we describe the analyzed data while in Section \ref{sec:Xray} we present the source identification and photometric analysis, along with spectroscopic and kinematic analysis. In Section \ref{Analisi A1} we present the detailed analysis of the X-ray AGN. We discuss our findings in Section \ref{Discussione}, and finally summarize our conclusions in Section \ref{sec:conclusioni}.

\section{Observation and Data Reduction}
\label{sec:obs and data}
\emph{\underline{HST images}}: HST images of \texttt{XDCP0044} have been obtained in 2015 with the Wide Field Camera 3 (WFC3) in F105W, F140W, F160W and F814W bands (Program: 13677, PI: S. Perlmutter) with the following exposure times: 4689 sec in F105W, 5189 sec in F140W, 2595 sec in F160W and 1620 sec in F814W. In our analysis, we use the images obtained by combining the archival drizzled (DRZ) frames, after performing the astrometry and aligning them. Due to the low S/N, the F814W band has not been included (see also \citealp{Travascio2020}). The depth of the three images at 5$\sigma$ 
is [28.53, 28.33, 28.18] respectively for F105W, F140W and F160W.

\emph{\underline{KMOS IFU data}}: observations in YJ (spectral coverage: 1.0251.344 $\mu$m) and H (spectral coverage: 1.4561.846 $\mu$m) bands have been obtained in 2013 (Program ID: 092.A0114(A), PI: R. Fassbender). Here we analyzed the KMOS data centered on the X-ray source 3 (RA=0:44:05.4570; DEC=-20:34:16.672) with a 2.8” $\times$ 2.8” field of view (see the magenta square in Figure \ref{fig:clusterreg3} \emph{right panel}), for a total target integration time of $\sim$4.4h in YJ band and $\sim$1.25h in H band.
At the redshift of the cluster (i.e. z$\sim$1.6), YJ-band samples the [OIII] and H$\beta$ emission lines with a resolving power R=3600 while H-band samples the H$\alpha$ region with R=4000. The data have been reduced using the pipeline with the Software Package for Astronomical Reduction with KMOS (SPARK; \citealp{Davies2013}), which includes dark correction, flat fielding, illumination correction, wavelength calibration and sky subtraction \citep{Davies2011}. The data are then combined according to the spatial shift of the objects in each frame. 

\emph{\underline{Chandra X-ray data}}: the data used for the X-ray spectral analysis are obtained from six different observations performed in September, October and November 2013 (PI: P. Tozzi) with the ACIS-S instrument of the \emph{Chandra} observatory for 370 ks of exposure time.
We consider X-ray data in the energy range 0.3-7.0 keV, where \emph{Chandra} ACIS detector is most sensitive and well calibrated. Details on observations and data reduction can be found in \citet{Tozzi2014}.

\section{Data Analysis}
\label{sec:Xray}

\subsection{NIR source identification and photometry}
\label{sub:identificazione e fotometria}
We focus our photometric analysis on the 24 kpc $\times$ 24 kpc region observed by HST (see \emph{right panel} of Figure \ref{fig:clusterreg3}).
Ground based (HAWK-I J and Ks bands and Subaru/Suprime V and i bands) photometric analysis in this region has been already performed by \citet{Fassbender2014} and \citet{Santos2014} who identified 3 different sources, i.e. A, C and D in Figure \ref{fig:clusterreg3} (\emph{right panel}), in addition to the previously discovered spectroscopic member B \citep{Santos2011}. 

Thanks to the higher resolution, HST images allow us to distinguish more sources compared to ground based images. We used SExtractor 
\citep{Bertin1996} to detect and deblend the sources, identifying a total of 9 objects.
As shown in Figure \ref{fig:clusterreg3} \emph{right panel}, A and B sources identified with HAWK-I consist of three different objects, i.e. \texttt{m1}+\texttt{m2}+\texttt{m3} and \texttt{n1}+\texttt{n2}+\texttt{n3}, respectively. Moreover, an additional faint source (\texttt{Q}) has been detected only in HST data. 

The photometric analysis has been performed using \textsc{a-phot} \citep{Merlin2019}. 
We tried to estimate the flux contamination from neighboring sources by 
taking two different photometric measurements. Photometric values are first obtained 
within elliptical apertures defined by eye to include most of the visible light coming from 
each galaxy (APER), which is then converted in magnitudes in the 
F105W, F140W and F160W bands using as zero points $zp=26.2,\, 26.4$ and $25.9$, 
respectively\footnote{See http://www.stsci.edu/hst/wfc3/ir\_phot\_zpt .}.  
A second photometric measurement is obtained simply
considering the fluxes assigned to each galaxy in the segmentation map 
created by SExtractor (ISO). We consider the difference between the two 
measurements as a reasonable approximation of the contamination between sources. 
Final uncertainties are computed as the combination of the statistical error 
provided by \textsc{a-phot} and the contamination uncertainty. 
We find that, given the extremely crowded region, the final errors 
are dominated by contamination. The derived HST F105W-, F140W- 
and F160W-band AB system magnitudes, together with the associated 
errors, and the HST coordinates (RA and DEC) are reported in Table \ref{table:mag}. \\

\begin{table*}
\caption{Identified sources in the high resolution HST images (see Figure \ref{fig:clusterreg3}, \emph{right panel}).}
\label{table:mag}      
\centering          
\begin{tabular}{c  c  c  c  c  c }     
\hline\hline       
HST ID & RA & DEC & $m_{F105W}$ & $m_{F140W}$ & $m_{F160W}$\\ 
\hline                    
    \texttt{m1}& 0:44:05.4570 & -20:34:16.672 & 23.54$\pm$1.32  & 22.61$\pm$1.06 &  22.17$\pm$0.92 \\  
    \texttt{m2}& 0:44:05.4369 & -20:34:15.962 & 23.91$\pm$0.06  & 23.38$\pm$0.01 &  23.13$\pm$0.01\\
    \texttt{m3}& 0:44:05.4008 & -20:34:16.470 & 23.75$\pm$0.01  & 23.06$\pm$0.13 &  22.81$\pm$0.17\\
    \hline
    \texttt{n1}& 0:44:05.5556 & -20:34:17.221 & 23.54$\pm$0.44  & 23.35$\pm$0.54 &  23.24$\pm$0.57\\
    \texttt{n2}& 0:44:05.5694 & -20:34:16.404 & 24.09$\pm$0.46 & 23.99$\pm$0.61 &  23.88$\pm$0.67\\
    \texttt{n3}& 0:44:05.5178 & -20:34:16.551 & 24.31$\pm$0.39  & 23.91$\pm$0.56 &  23.69$\pm$0.63\\
    \hline
    \texttt{O}& 0:44:05.5080 & -20:34:15.348  & 23.08$\pm$0.37  & 22.37$\pm$0.32 & 22.21$\pm$0.32 \\
    \hline
    \texttt{P}& 0:44:05.4300 & -20:34:14.715  & 23.12$\pm$0.36  & 22.55$\pm$0.37 & 22.45$\pm$0.39  \\
    \hline
    \texttt{Q}& 0:44:05.5050 & -20:34:17.968  & 24.75$\pm$1.14  & 24.19$\pm$1.05 & 24.06$\pm$1.09\\
\hline                  
\end{tabular}
\tablefoot{Columns are: HST ID and position (RA and DEC), F105W-, F140W- and F160W-band 
magnitudes with their uncertainties including the statistical error computed 
by \textsc{a-phot} and the contamination uncertainty.}
\end{table*}

\subsection{NIR spectroscopy}
\label{sub:spettroscopia NIR}

\begin{figure*}
 \centering
            {\includegraphics[scale=0.35]{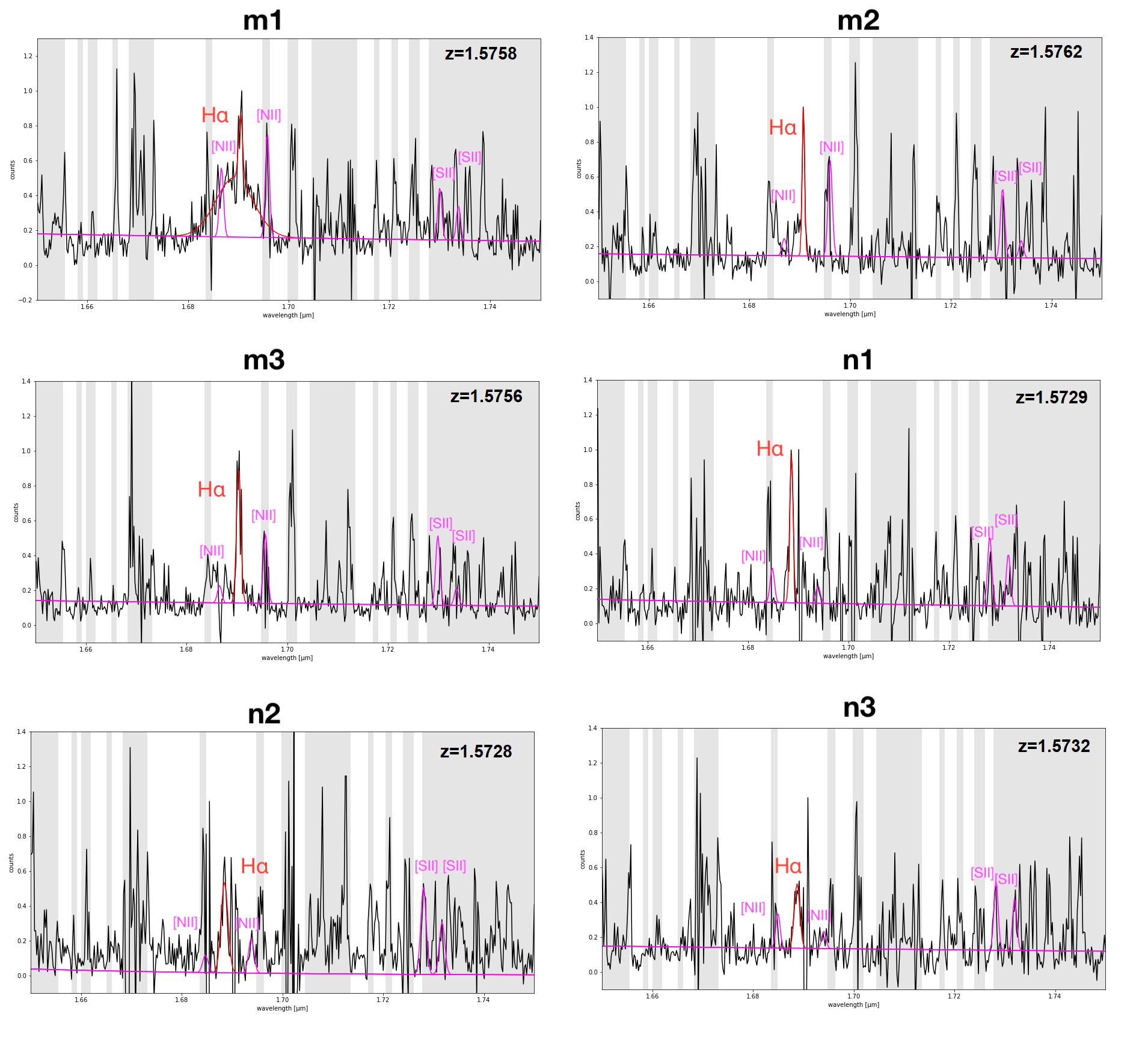}}
      \caption{Zoom-in of the H-band spectra around the H$\alpha$ line of the 6 confirmed cluster members. In red it is shown the power-law+gaussian fit of the emission line, while the magenta line is used to mark the continuum and the [NII] and [SII] emission lines. The derived spectral properties are listed in Table \ref{table:fit spettroscopia}.} 
         \label{fig:fit spettri}
   \end{figure*}

Here we present the spectroscopic analysis of the sources identified in HST. 
Spectra in YJ- and H-band have been extracted on the position of the sources identified in HST, using a fixed aperture of 3 pixel (i.e. 0.6'') in diameter on the KMOS data cubes.
In the H-band spectrum of 6 out of the 9 sources we found individual emission lines with SNR ranging from $\sim 3$ to $\sim 10$. As these sources belong to a crowded region, it is reasonable to assume they are physically close to each other, i.e. having similar redshift. Moreover, since these complexes are within a galaxy cluster, we can guess that the redshift of the galaxies is approximately similar to the average cluster redshift. Thus, we assume that these lines represent H$\alpha$ transitions, 
which is also one of the strongest line expected.
For two of them (\texttt{m1} and \texttt{n1}),  an H$\beta \lambda 4861\text{\AA}$ line is also observed in the YJ-band spectrum, although we found these lines into a noisy region contaminated by the sky lines.
A zoom-in of the spectral region showing the H$\alpha$ line for these sources is shown in Figure \ref{fig:fit spettri}, while the redshifts derived by fitting the observed lines are listed in Table \ref{table:fit spettroscopia} together with the FWHM of the line and the SNR calculated as the ratio between the signal at the peak of the narrow H$\alpha$ emission line and the $\sigma$ of the signal in the spectrum where there is only the continuum.

\begin{table*}
\caption{Results of the KMOS spectral analysis.}
\label{table:fit spettroscopia}      
\centering          
\begin{tabular}{c  c  c  c  c  c  c  c}     
\hline\hline       
ID & z & $\rm FWHM_{B}$ & $\rm Log(L_{H\alpha B}$) & $\rm FWHM_{N}$& $\rm Log(L_{5100\text{\AA}}$) & $M_{V}$ & SNR \\ 
   &   & km/s       & erg/s &  km/s      & erg/s              & &         \\
\hline                    
    \texttt{m1}& 1.5758 & 1551$\pm$102 & $\sim$42.84 & 167$\pm$30 & 43.85$\pm$0.21 & -22.39$\pm$0.09 & 3.88\\  
    \texttt{m2}& 1.5762 &            &  & 177$\pm$19 & 43.41$\pm$0.67 & -21.17$\pm$0.32 & 6.14\\
    \texttt{m3}& 1.5756 &            &  & 177$\pm$14 & 43.58$\pm$0.59 &  -21.64$\pm$0.28 & 8.24\\
    \hline
    \texttt{n1}& 1.5729 &            &  & 177$\pm$12 & 43.47$\pm$0.73 &  -21.22$\pm$0.35 & 9.85\\
    \texttt{n2}& 1.5728 &            &  & 226$\pm$27 & 43.32$\pm$0.33 &  -20.82$\pm$0.16 & 4.95\\
    \texttt{n3}& 1.5732* &           &  & 176$\pm$42 & 43.23$\pm$0.36 &  -20.67$\pm$0.16 & 2.79\\
\hline                  
\end{tabular}
\tablefoot{(1) source ID; (2) redshift; (3) FWHM of the H$\alpha$ emission line (broadB component); (4) Luminosity of the broad H$\alpha$ emission line; (5) FWHM of the H$\alpha$ emission line (narrowN component);  (6) 5100$\text{\AA}$ Luminosity; (7) Vband absolute magnitude; (8) SNR ratio.}
\end{table*}

\begin{figure*}
 \centering
           {\includegraphics[scale=0.25]{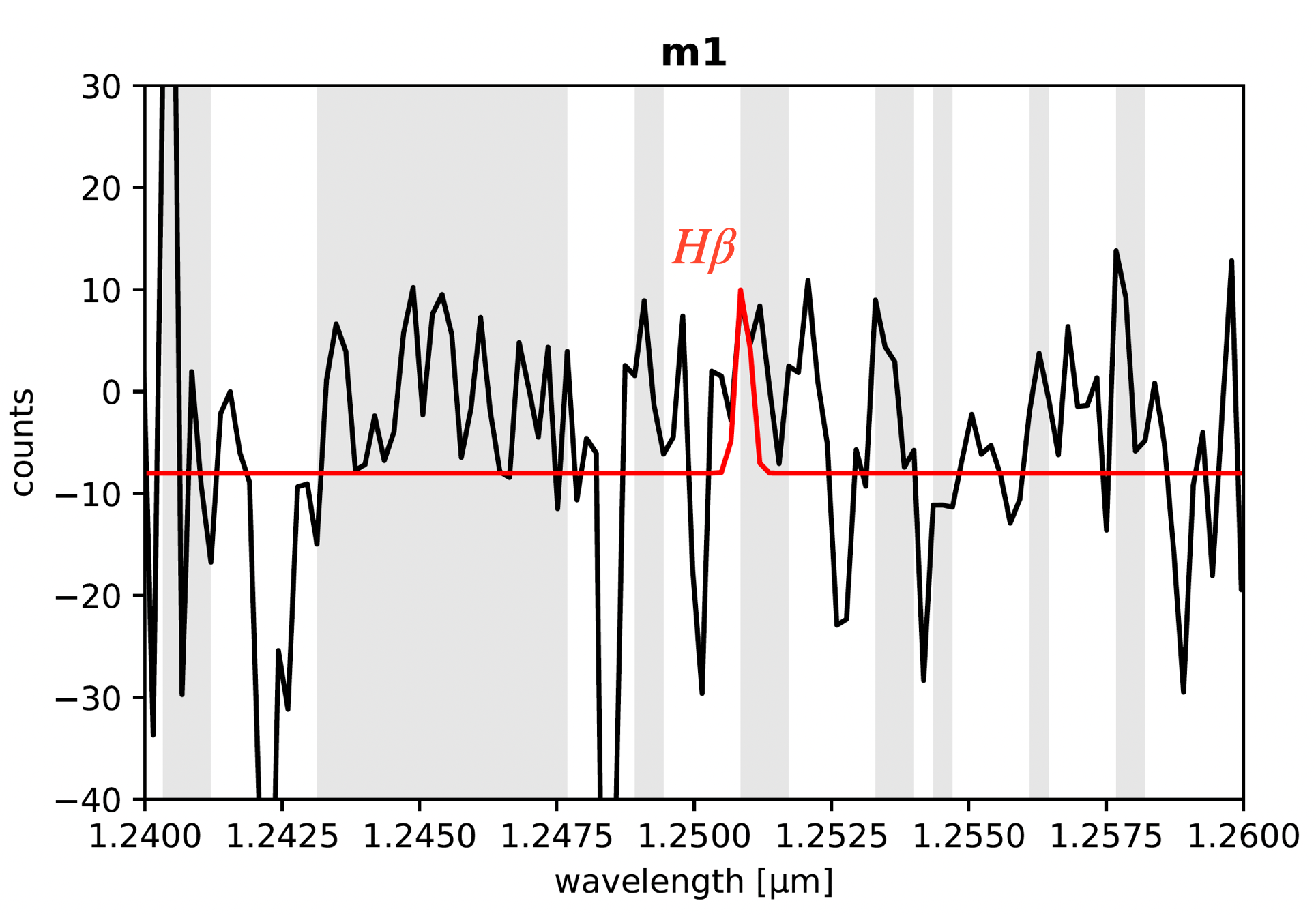}
           \includegraphics[scale=0.25]{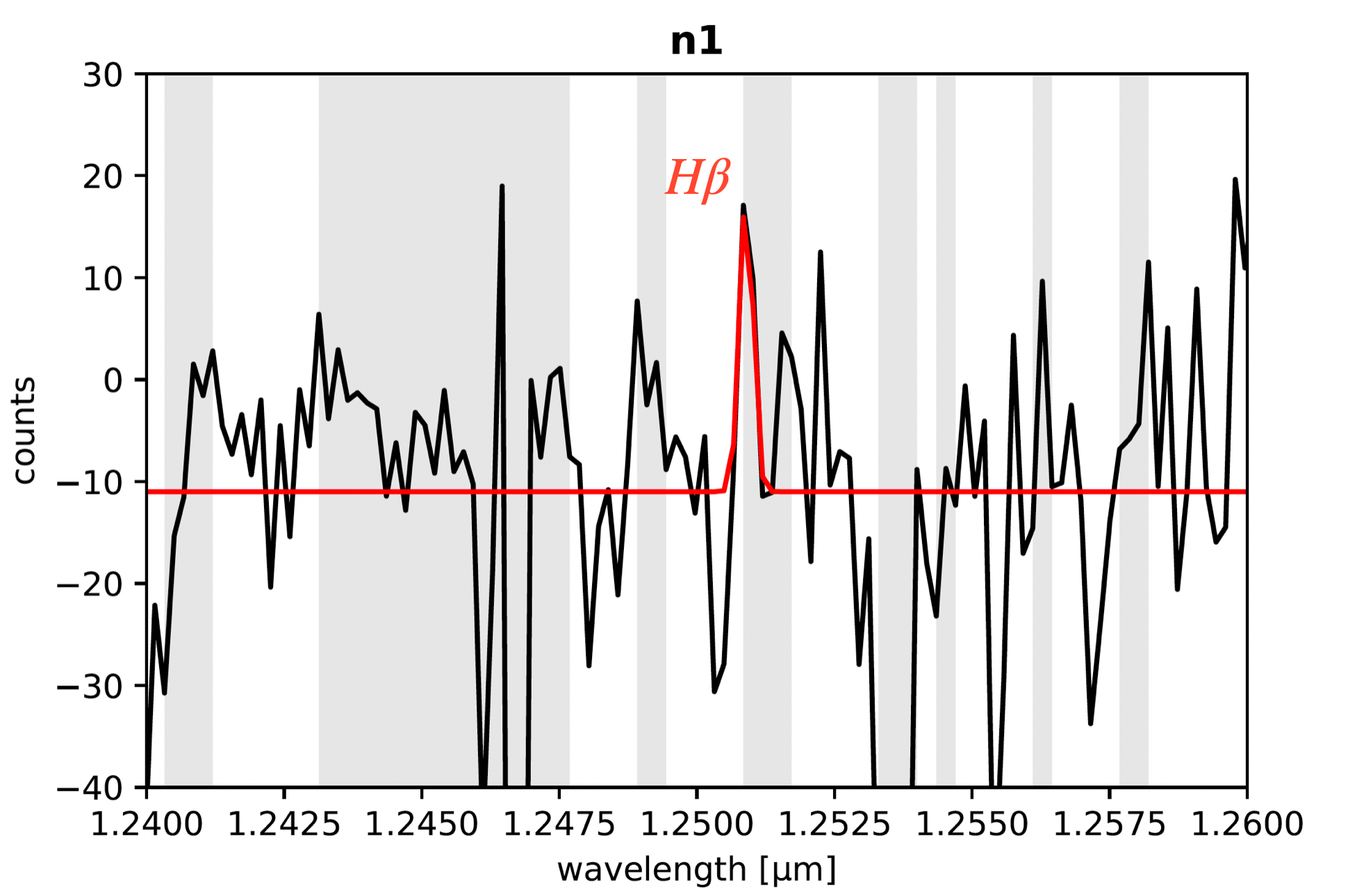}}
            \caption{Zoom-in of the YJ-band spectra around the H$\beta$ line respectively for source \texttt{m1} and \texttt{n1}.}
         \label{fig:spettri H beta}
\end{figure*}

Redshifts range from $z=1.5728$ to $z=1.5762$ and, given the small dispersion ($\Delta z=0.0034$), all 6 galaxies are therefore spectroscopically confirmed cluster members. Given the redshifts of the sources and their apparent position, we can identify in the system two subgroups: \texttt{Complex M}, which includes \texttt{m1}, \texttt{m2} and \texttt{m3} whose redshifts range from $z=1.5756$ to $z=1.5762$ 
($\Delta z=0.0006$) and \texttt{Complex N}, consisting of \texttt{n1}, \texttt{n2} and \texttt{n3} with z between 1.5728 and 1.5732 ($\Delta z=0.0004$). Note that the redshift of source \texttt{n3} is tentative given the low Signal-to-Noise Ratio (SNR) of the line. 
Interestingly, the analysis of the spectrum of source \texttt{m1} shows that the H$\alpha$ emission line is broad ($FWHM>1500$ km/s, see Figure \ref{fig:fit spettri} \texttt{m1}-panel). Source \texttt{m1} is therefore classified as a broad line AGN (BLAGN, see Section \ref{Analisi A1} for more details). Finally, from the analysis of the YJ-band spectra, an H$\beta \lambda 4861\text{\AA}$ line is observed for sources \texttt{m1} and \texttt{n1},  confirming their redshift (see Figure \ref{fig:spettri H beta}). 

For all 6 galaxies for which a redshift has been measured, we have estimated the rest-frame luminosity at 5100$\text{\AA}$ ($L_{5100\text{\AA}}$) and the V-band absolute magnitude from the flux measured at the wavelength of interest, by interpolating the HST
photometric points in F105W and F140W bands. These values are listed in Table \ref{table:fit spettroscopia}.

\subsection{Kinematic analysis of the super complex of galaxies}
\label{sub:mappe}
\begin{figure}
 \centering
            {\includegraphics[scale=0.4]{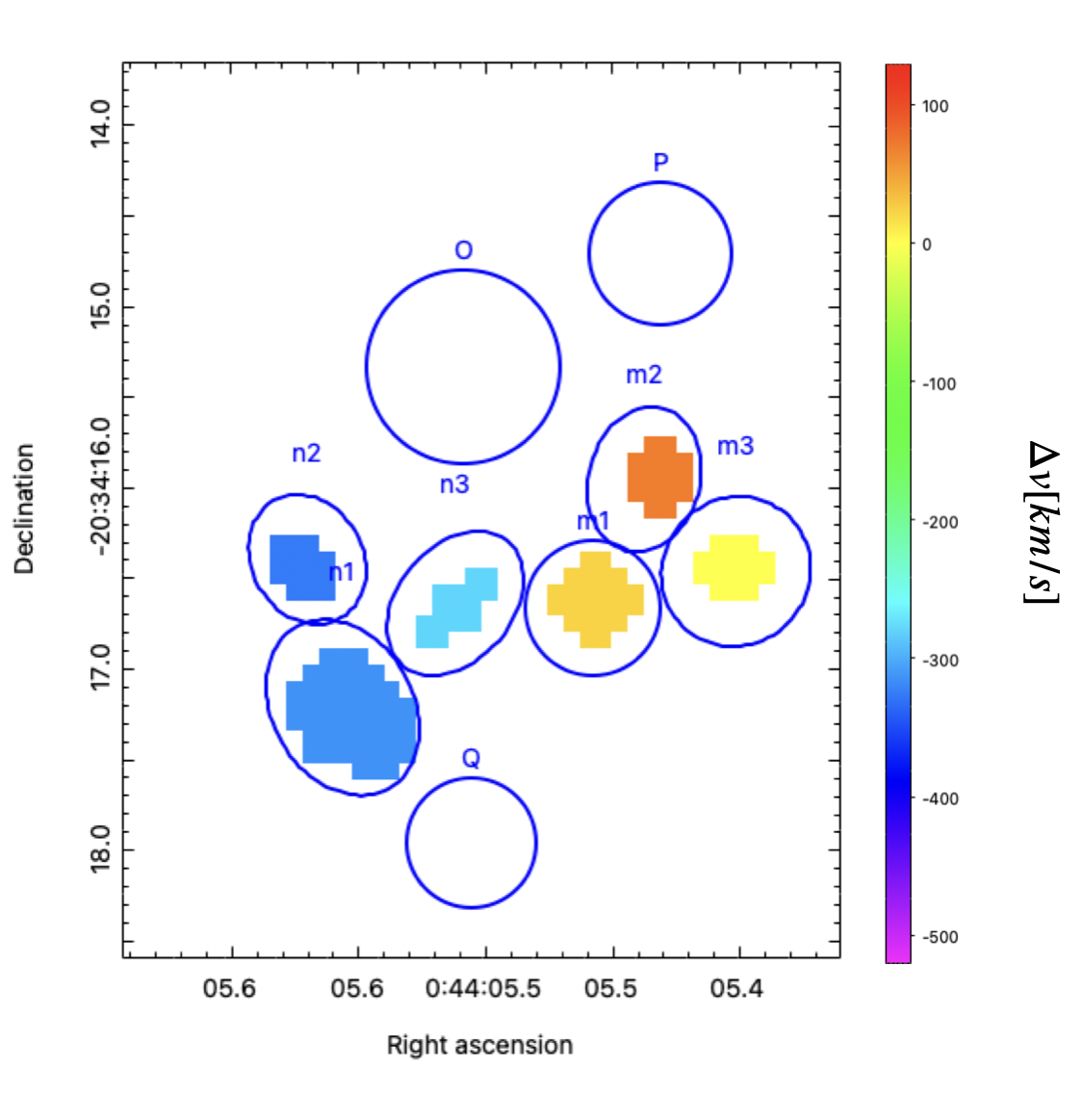}}
            \caption{Velocity shift map of the individual sources, with respect to \texttt{m1}. Blue circles mark the sources detected in HST
            .}
         \label{fig:vel shift sorgenti}
   \end{figure}

\begin{figure}
 \centering
            {\includegraphics[scale=0.45]{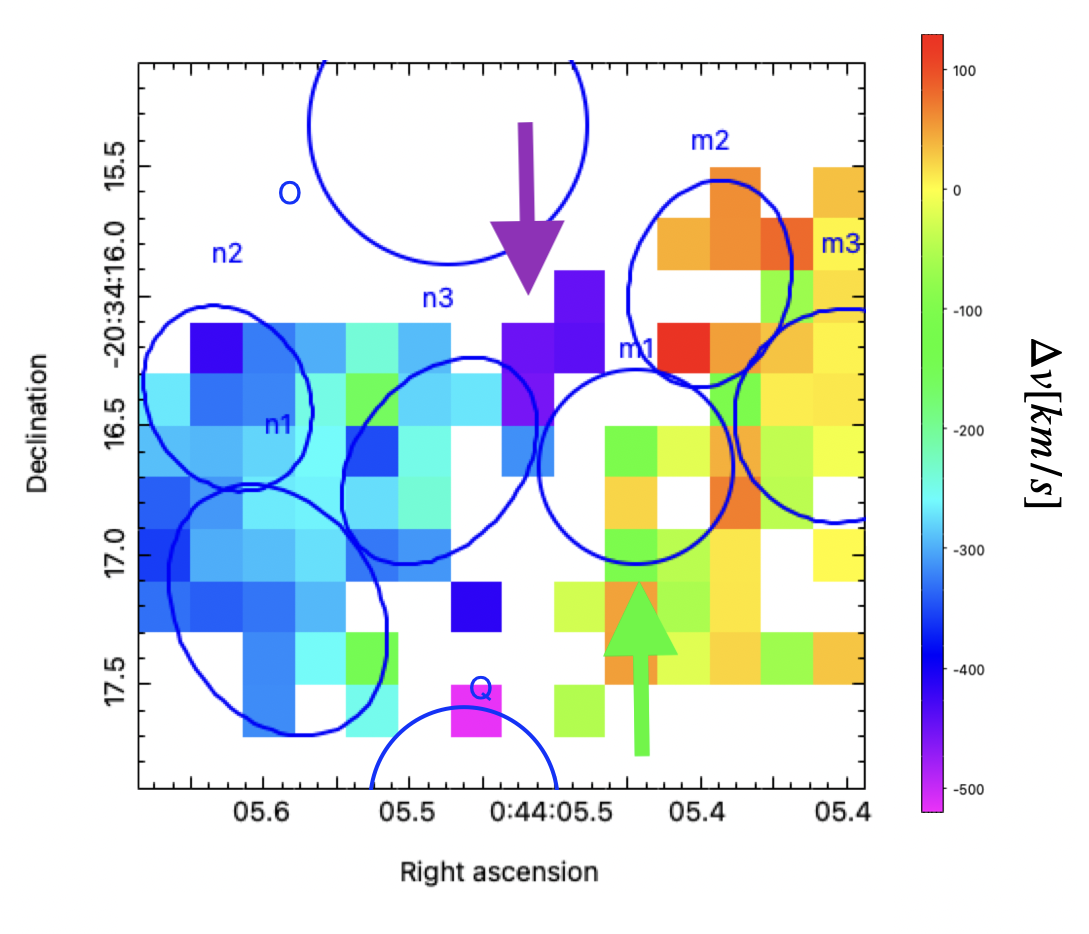}}
            \caption{Velocity shift map of the narrow H$\alpha$ emission line. Green and purple arrows mark the H$\alpha$ extended emission detected outside the main sources.}
         \label{fig:vel shift completa}
   \end{figure}

\texttt{XDCP0044} shows a complex dynamical state. One hint of this particular state is the presence of diffuse emission between galaxies. Therefore, we search for emission lines at the same redshift of the analyzed system. Particularly, the IFU data allow us to perform a spatially resolved study of the kinematics of the H$\alpha$ emission line. \\
As a first step, we create a velocity shift map of the narrow H$\alpha$ emission line relative to source \texttt{m1} to identify patterns of gas and galaxies. We choose source \texttt{m1} because it is an AGN (see Section \ref{Analisi A1}), therefore, is the best candidate to be the most massive galaxy 
and the gravitational center of the system.
We estimated the velocity shifts for each galaxy considering source \texttt{m1} as reference and following the relation by \cite{Harrison1974}: 
\begin{equation}
    v_{shift}=\biggr(\frac{\lambda_{obs}-\lambda_{cen}}{\lambda_{cen}}\biggr)\, c \, ,
\end{equation}
where $\lambda_{obs}$ is the observed wavelength of the narrow H$\alpha$ emission line in a given source, while $\lambda_{cen}$ is the wavelength corresponding to the H$\alpha$ emission line of source \texttt{m1}.
Figure \ref{fig:vel shift sorgenti} shows a map in which  different colors indicate the velocity shift of each source with respect to source \texttt{m1}. This map has been obtained extracting the total emission of narrow and broad H$\alpha$ emission lines for each source considering the regions used for photometric analysis (see Section \ref{sub:identificazione e fotometria}). We then studied the dynamics of the diffuse emission by looking at the narrow H$\alpha$ emission line pixel by pixel, excluding the broad H$\alpha$ emission line (see Figure \ref{fig:vel shift completa}). The SNR cut of these map is $\sim$ 2.5.

A visual inspection of the maps clearly confirm that sources
\texttt{m1}, \texttt{m2} and \texttt{m3} belong to the same complex of galaxies (\texttt{Complex M}), while sources \texttt{n1}, \texttt{n2} and \texttt{n3} to \texttt{Complex N}, with a negative velocity shift ($\sim$ $300$ km/s) with respect to the \texttt{Complex M}. 
Given the small velocity difference, \texttt{Complex M} and \texttt{Complex N} are expected to merge individually to form two more massive galaxies. Eventually, \texttt{Complex M} and \texttt{Complex N} may also merge with each other, finally forming a second BCG, as anticipated by the analysis of \cite{Fassbender2014}. These results will be further discussed in Section \ref{Mass assembly}.

Then, we focus on a region located south of source \texttt{m1}, where we identify  diffuse emission at the same wavelength of the narrow H$\alpha$ emission line of \texttt{m1}, which is also visible in the HST IR images. In addition, we find diffuse emission also between sources \texttt{m1} and \texttt{n3}, with a very high velocity shift ($\sim$ $ 450 km/s$), although the detection has a low significance. 
Both detections can be interpreted as diffuse gas in the midst of galaxies, destabilized by recent galaxy-galaxy encounters. In the case of the H$\alpha$ emission below source \texttt{m1}, this could also be interpreted as extended ($\sim$ 7 kpc) Narrow Line Region (ENLR) around source \texttt{m1} (see e.g. \citealp{Hippelein1996}, \citealp{Husemann2019}). We will not discuss further the diffuse emission, arguing that deeper IR integral field spectroscopy may reveal in greater detail the dynamics of the diffuse baryons in the dense cluster core, 
helping in tracing the complex dynamical interaction occurring during the merging phase.

\section{Analysis of the source \texttt{m1}}
\label{Analisi A1}

\subsection{X-ray spectroscopy of \texttt{m1}}
\label{sub:Xray spectroscopy}

\begin{figure}
\centering
\includegraphics[scale=0.4]{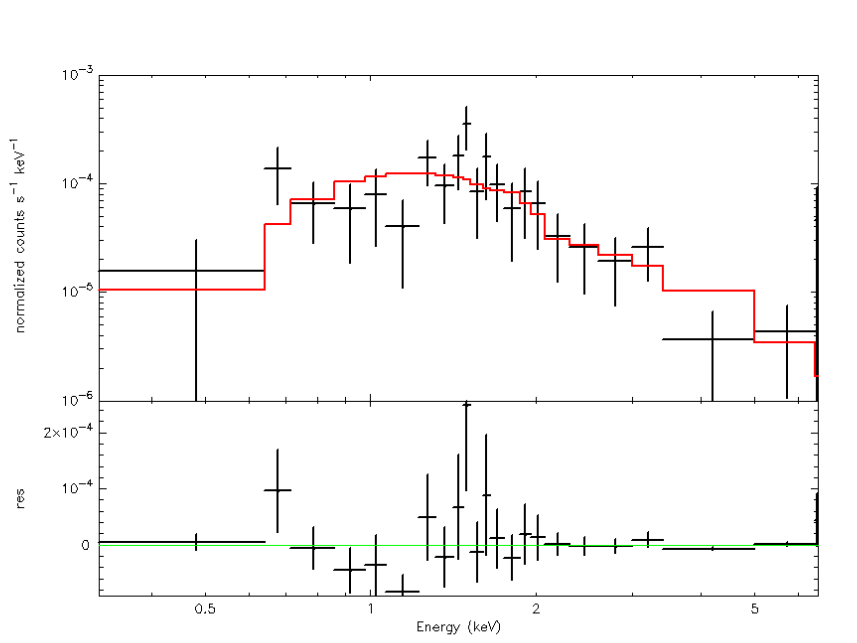}
\caption{\emph{Chandra} X-ray spectrum of \texttt{m1} with the best-fit model (red line). Lower panel shows the residuals.}
\label{fig:spettro X unico}
\end{figure}

As discussed in Section \ref{sub:spettroscopia NIR}, the source \texttt{m1} is identified as a BLAGN and has been detected as a point-like source in the \emph{Chandra} image by \cite{Tozzi2014}. 
Here we present a detailed spectral analysis of the \emph{Chandra} data of \texttt{m1}. The source+background spectrum has been extracted from a circular region of radius $\sim$ 1 arcsec 
centered on source \texttt{m1}, within which we measure $72.8^{+9.6}_{-8.5}$
net counts in the 0.3-7 keV band. The background spectrum (rescaled by the extraction area and subtracted to the source+background spectrum) has been extracted from an annular region of inner and outer radii 4 and 10 arcsec, respectively. 
The data from these six observations (see Section \ref{sec:obs and data}) are combined to form a single spectrum using the FTOOLS script \texttt{addascaspec}. Then, the spectra are binned so that there was at least 1 count per bin and modeled using the software XSPEC v. 12.12.0 in the [0.3-7.0 keV] band. Also, we performed the analysis using Cash statistics \citep{Cash1979}. 
The rebinned spectrum is shown in Figure \ref{fig:spettro X unico}.

To fit the spectrum we used a model that includes three terms: a power law which models the emission related to the Comptonization of photons from the hot electron corona surrounding the accretion disk, a term which accounts for the Galactic absorption and a term for the intrinsic absorption. The results show that the photon index $\Gamma$ is between 1.8 and 2.0, the typical value for Type1 AGNs, and that the source is moderately obscured ($\rm Log[N_{H}/cm^{2}]=22.4^{+0.2}_{-0.5}$), assuming a Galactic column density fixed at $\rm Log[N_{H}/cm^{2}] = 20.1$ \citep{HI4PI}. The X-ray properties of the source \texttt{m1} are reported in Table \ref{table:prop A1}. 
From the derived values, and assuming a redshift of 1.5758 (see Section \ref{sub:spettroscopia NIR}), the intrinsic, unabsorbed luminosity is measured to be $\rm Log[L_{[2-10 keV]}^{X}]= 43.4^{+0.3}_{-0.4}$ in the [2-10 keV] rest-frame band.
These results are consistent within $1\sigma$ with those 
found in \cite{Tozzi2014}, where an aperture of 1.5 arcsec is used for the extraction of the spectrum.

In order to test possible variability in flux, slope and absorption during the three months period of observations, we considered the three spectra separately obtained by merging the corresponding Obsid in three different periods, and jointly modeled them. In this case, the model used to fit the spectra is the same as in the case of a single spectrum with the addition of a constant that takes into account possible changes in the source flux.  We find that the signal in each temporal bin is too low to search for 
spectral variability, while, focusing on the flux normalization, we found no hints for variability.

\subsection{Bolometric luminosity, BH mass, and Eddington ratio}
\label{sub:MBH lambda LEdd}
From the $5100 \text{\AA}$ luminosity, we derived the bolometric luminosity using the relation
\begin{equation}
    K_{5100\text{\AA}}=\frac{L_{bol}}{L_{5100\text{\AA}}}
\end{equation}
where $K_{5100\text{\AA}}$ is the bolometric correction given by \cite{Krawczyk2013} (see also \citealp{Saccheo2022}). Considering $K_{5100\text{\AA}}=4.33\pm1.29$, the value obtained for source \texttt{m1} is $\rm Log[L_{bol}/erg/s]=44.49^{+0.11}_{-0.15}$. \\
It is also possible to compute the bolometric luminosity using the X-ray luminosity at [2-10 keV] band and assuming the bolometric corrections by \cite{Duras2020}:
\begin{equation}
   \rm  K_{X}(Bol)=\frac{L_{bol}}{L_{X}}=a \biggr[1+\biggr(\frac{log(L_{bol}/L_{\odot})}{b}\biggr)^{c}\biggr]
\end{equation}
with $a=12.76 \pm 0.13$, $b=12.15 \pm 0.01$ and $c=18.78\pm 0.14$ as in Table 1 of \cite{Duras2020} for Type1 AGN. The value obtained for source \texttt{m1} is $\rm Log[L_{bol}^{X}/erg/s]\sim 44.52$, in agreement 
to the one obtained from the 5100$\text{\AA}$ luminosity above. Given the nuclear origin of the AGN X-ray emission, such agreement ensures us that the 5100$\text{\AA}$ luminosity is not contaminated by the host galaxy.  

We then estimated the BH mass of \texttt{m1} using the virial formula applied to the Broad Line Regions (BLRs). In particular, since in the spectrum of \texttt{m1} we detected a broad H$\alpha$ emission line, we used the relation found by \cite{Greene_Ho2005} which relates to the FWHM and the luminosity of the broad H$\alpha$ emission line
according to the relation:
\begin{equation}
    \rm M_{BH}=2.0^{+0.4}_{-0.3}\times 10^{6} \biggr(\frac{L_{H\alpha}}{10^{42} {\rm erg/s}}\biggr)^{0.55\pm0.02} \biggr(\frac{FWHM_{H\alpha}}{10^{3} {\rm km/s}}\biggr)^{2.06\pm0.06} M_{\odot}\, .
\end{equation}

\noindent
Using the values of $\rm FWHM_{H\alpha}$ and $\rm L_{H\alpha}$ from Table \ref{table:fit spettroscopia}, we derived the mass of the central BH which is $\rm Log[M_{BH}/M_{\odot}]= 7.15\pm0.32$. 
As a further check, we computed the BH mass using, together with the FWHM of the broad H$\alpha$ emission line, the absorption corrected [2-10 keV] X-ray luminosity as in Eq. (4) of \cite{Bongiorno2014}:
\begin{equation}
    log M_{BH}=7.11+2.06log\frac{FWHM_{H\alpha}}{10^{3}{\rm km/s}}+0.693log\frac{L_{[2-10 {\rm keV}]}}{10^{44}{\rm erg/s}}\, .
\end{equation}
The obtained value is $\rm Log[M_{BH}/M_{\odot}]= 7.08^{+0.40}_{-0.45}$, consistent with the value found above. \\
The error associated to the BH masses are  given by the sum of the statistical and systematic uncertainties. The systematic uncertainty in the $\rm LogM_{BH}$ determination has been estimated in 0.3 dex to account for the observed scatter in the virial relation itself, while in the computation of the statistical errors, we take into account the errors in the 5100\text{\AA} and X-ray luminosity, and the one on the FWHM
measurement (in quadrature).\\
Furthermore, from the central BH mass it is possible to calculate the Eddington luminosity which is linked only to the mass of the body that is accreting according to the relation
\begin{equation}
   \rm  L_{Edd}=1.33 \times 10^{38} \biggr(\frac{M}{M_{\odot}}\biggr) \, [erg/s].
\end{equation}
Assuming $\rm Log[M_{BH}/M_{\odot}]= 7.15\pm0.32$ ($7.08^{+0.40}_{-0.45}$), the Eddington luminosity of source \texttt{m1} is $\rm Log[L_{Edd}/erg/s]=45.28\pm0.32$ ($45.21^{+0.40}_{-0.45}$). \\
Finally, we calculated the Eddington ratio
\begin{equation}
    \lambda_{Edd}=\frac{L_{bol}}{L_{Edd}}
\end{equation}
which describes the rate at which the central supermassive (SM) BH is accreting. We obtain a value in 
the range $\lambda_{Edd}\sim$0.17-0.20.
Therefore, we conclude that source \texttt{m1} hosts a relatively small ($\sim 10^{7} M_{\odot}$) and moderately accreting ($\lambda_{Edd} \sim 0.2$) SMBH. 
Its physical properties are summarized in Table \ref{table:prop A1}.

\begin{table*}
\caption{Physical properties of the source \texttt{m1}.}
\label{table:prop A1}      
\centering          
\begin{tabular}{c  c  c  c  c  c  c c}     
\hline
ID & $\rm Log[L_{bol}]$ & $\rm Log[M_{BH}]$ & $\lambda_{Edd}$ & $\rm Log[L_{X}]$ & $\rm Log[N_{H}]$ & $\Gamma$  \\ 
   &  erg/s         & $M_{\odot}$  &                 & erg/s & $cm^{-2}$   &            \\
\hline
&&&&&& \\
\texttt{m1} & $44.49 ^{+0.11}_{-0.15}$   & $7.15\pm0.32$ & $\sim$0.17-0.20 & $43.4^{+0.3}_{-0.4}$ & $22.4^{+0.2}_{-0.5}$ & $1.9^{+0.5}_{-0.4}$ \\
&&&&&& \\
\hline                    
\end{tabular} 
\tablefoot{$\rm Log[L_{bol}]$ refers to the bolometric luminosity computed from $\rm L_{[5100\text{\AA}]}$ with the bolometric correction by \cite{Krawczyk2013}.}
\end{table*}

\section{Discussion}  
\label{Discussione}

\subsection{Mass assembly and time scales}
\label{Mass assembly}
Models and simulations \citep{DeLucia2007} predict that the mass of the galaxy population observed in clusters in the local Universe began to assemble at $z\sim 35$. At $z\sim 2.5$ the progenitors of massive galaxies start to form and at $z \sim 12$ merger activity reaches a peak. Then, the mass assembly proceeds through minor merging processes, leading to the formation of galaxies observed in the local Universe. \\
As we previously mentioned, $z\sim 1.6$ is a crucial epoch to study the mass assembly because it is close to the SF and BH activity peak \citep{Madau2014}.
The region of 24 kpc $\times$ 24 kpc analyzed in this work 
includes a super complex formed by 6 up to 9 interacting galaxies, making it 
an extraordinarily dense region. The surface number density is measured to be $\sim10^{-2}$ kpc$^{-2}$ in physical units. This can be seen as a lower limit, considering we find 9 sources identified in HST images. This region is denser than the core region of \texttt{XDCP0044} \citep{Travascio2020}, which shows a surface number density one order of magnitude lower and also even denser than the core region of the Spiderweb protocluster at $z\sim 2.156$ (\citealp{Roettgering1994}, \citealp{Pentericci1997}, \citealp{Miley2006}), which shows a surface number density of $\sim 7 \times 10^{-4}$ kpc$^{-2}$ \citep{Kuiper2011}. 

Furthermore, the 6 confirmed cluster members form two different complexes, \texttt{Complex M} and \texttt{Complex N}, which are located at a distance $d \sim 13$ kpc and have a relative velocity $v\sim 300-400$ km/s. Simulations show that most pairs of galaxies near enough to each other, i.e. typically 20-30 kpc, and with a low velocity difference ($\sim$ 200-300 km/s) will eventually merge on a short time scale (e.g. \citealp{Patton2002}, \citealp{Lopez-Sanjuan2011}, \citealp{Lopez-Sanjuan2012}). Indeed, the 6 to 9 interacting galaxies in the analyzed region will likely merge 
within a timescale comparable to dynamical friction \citep{Conselice2014}. \\
It is also possible to estimate the collision timescale among sources in \texttt{Complex M} and sources in \texttt{Complex N}, considered pairs of galaxies in a dense environment. 
To estimate this timescale we first computed the stellar mass of each galaxy resolved by HST images by performing SED-fitting procedure on HST photometry, HAWK-I J and Ks bands, Subaru/Suprime V and i bands. We used the Z-PHOT code \citep{Fontana2000}, with \cite{BruzualCharlot2003} templates, \cite{Salpeter1955} initial mass function and \cite{Calzetti2000} extinction. We adopted exponentially declining SF histories ($\tau$models).  
However, for the detected sources \texttt{n1}, \texttt{n2} and \texttt{n3}, the accuracy on photometry is not sufficient to constrain the SED, thus masses have been derived assuming different values of mass-to-light ratio, according to different population models \citep{McGaugh2014}. Mass values
and associated uncertainties are listed in Table \ref{table:masse}. 
Finally, we estimated the radius of the sources as the half-light radius obtained with SExtractor \citep{Bertin1996}. 
The best-fitting masses of the sources in the analyzed region sum up to a total stellar mass of $\sim 10^{11} M_{\odot}$.

\begin{table*}
\caption{Values of mass, with upper and lower limits, for the six sources in the KMOS field of view.}
\label{table:masse}      
\centering          
\begin{tabular}{c  c  c c}     
ID & Log[M] & Log[$M_{Max}$] & Log[$M_{Min}$]  \\ 
   &  $M_{\odot}$         &   $M_{\odot}$ & $M_{\odot}$\\
\hline
\texttt{m1} & 9.4 & 10.9 & 8.6 \\
\texttt{m2} & 10.1 & 10.3 & 8.8 \\
\texttt{m3} & 8.5 & 9.8 & 8.3 \\
\hline
\texttt{n1} & 9.4* & 10.1* & 8.9* \\
\texttt{n2} & 9.2* & 9.9* & 8.8* \\
\texttt{n3} & 9.2* & 9.8* & 8.8* \\
\hline
\end{tabular}
\tablefoot{The values marked with "*" are those calculated assuming different mass-to-light ratios from \cite{McGaugh2014}.}
\end{table*}

Then, we used the relation reported by \cite{Vijayaraghavan_Ricker2013} 
to calculate the collision timescale: 
\begin{equation}
    t_{coll}=\frac{1}{n_{gal}\sigma_{cs}v_{gal}}\, ,
\end{equation}
where
\begin{itemize}
    \item $n_{gal}$ is the number density of galaxies in the analyzed region. In our case we considered the cluster members detected in the KMOS field of view in a sphere with radius of $\sim 12$ kpc (half of the KMOS field of view);
    \item $\sigma_{cs}$ is the galaxy cross-section calculated as
    \begin{equation}
        \sigma_{cs}\sim\pi(r_{1}^{2}+r_{2}^{2})\biggr(1+\frac{G(M_{1}+M_{2})}{(r_{1}+r_{2})v_{rel}^{2}}\biggr)\biggr(\frac{v_{esc}}{v_{gal}}\biggr)^{2/3}  \, ,
    \end{equation}
    with $M_{i}$ and $r_{i}$ are the mass and radius of the galaxy pair and $v_{esc}$ the escape velocity from the system;
    \item $v_{gal}$ is the velocity shift between the sources.
\end{itemize}
The resulting merging time is of the order of $\sim$ 10 Myr for sources in \texttt{Complex M} and $\sim$ 30 Myr for sources in \texttt{Complex N}.
If we use the same approach to estimate the merging time between
\texttt{Complex M} and \texttt{Complex N} we obtain an estimate of 
$\sim$ 370 Myr for this two complexes to form a single source.

\subsection{Nuclear and star formation activity}
\label{AGN e SF activity}

One of the results of this work is the presence of an AGN 
in a very dense ($\sim 4.9 \times 10^{8} M_{\odot}/$kpc$^{2}$) 
and star-forming (SFR=825$\pm$120 $M_{\odot}/yr$, see Section \ref{sec:Intro}) region 
at $\sim 157$ kpc from core of a galaxy cluster at $z>1.5$. 
The properties of this AGN are relevant to the AGN-host galaxy co-evolution scenario. 
According to \cite{Martini2009}, \cite{Alberts2016}, \cite{Bufanda2017} and other studies, the presence of an AGN in a galaxy cluster is directly correlated with the presence of gas-rich galaxies that, through merging, influence the transportation of molecular gas and consequently the AGN activity. The high SFR 
measured implies a large quantity of molecular gas surrounding the system. The availability of this in-flowing gas means, in turn, that there are conditions for triggering BH accretion and, at the same time, sustain the vigorous SF itself. Clearly, the dense environment favour SFG mergers and thus again accelerate both SF and nuclear activity \citep{Ellingson1993}.  \\ 
The AGN at the center of the studied region appears to be moderately obscured ($\rm Log[N_{H}/cm^{2}]\sim 22$), hosting a medium size ($M_{BH}\sim10^{7}M_{\odot}$) and moderately accreting ($\lambda_{Edd}\sim0.2$) BH (see Section 4.1), providing a luminosity $\rm Log[L_{X}/erg/s]\sim 43$.
Following \cite{Sanders1988} and \cite{Hopkins2006} evolutionary models, this represents a specific phase in the AGN-galaxy co-evolution. In these models gas cools during galaxy mergers and fall onto the central BH, feeding its accretion and SF in the host galaxy. During this phase, the central BH is obscured by gas and dust. After all the gas is consumed or displaced by nuclear and stellar feedback, the AGN luminosity and star formation episodes decrease. Particularly, the AGN studied in this work may be in the early stages of accretion.
During this phase we find a high SFR, due to the presence of cold gas, which is feeding also the central BH.
According to this scenario, the AGN will continue to accrete gas from the environment/host galaxy, increasing its mass until, at some point, it will reach a critical mass and enter the feedback phase, beginning to emit winds and outflows. These winds and outflows eventually quench the SF in the host galaxy, 
driving it toward a red and passive galaxy, i.e. a possible second BCG located south of the cluster core.

\subsection{\texttt{XDCP0044}: a double merging cluster}

\begin{figure*}
 \centering
           {\includegraphics[scale=0.3]{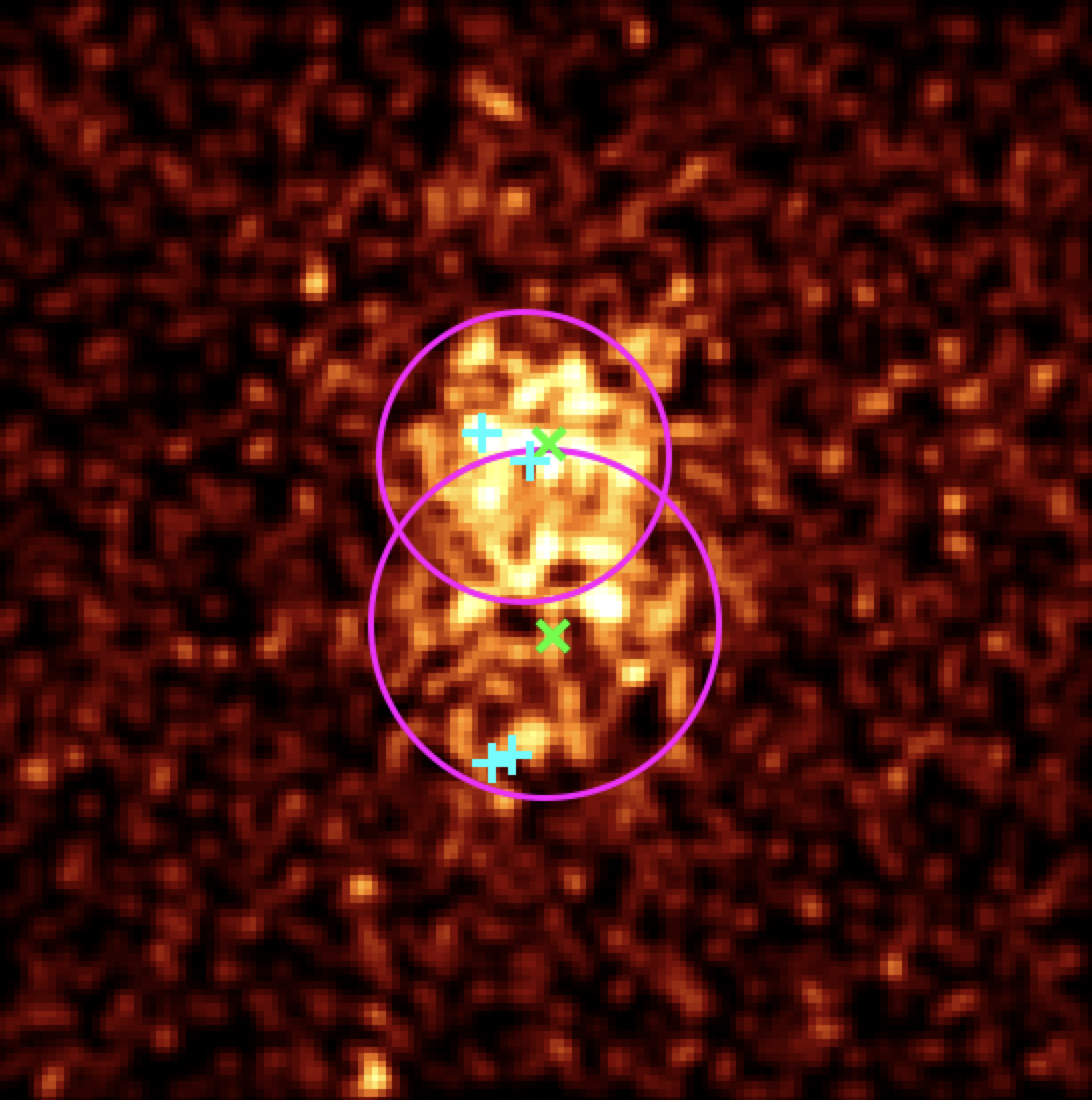}
           \includegraphics[scale=0.3]{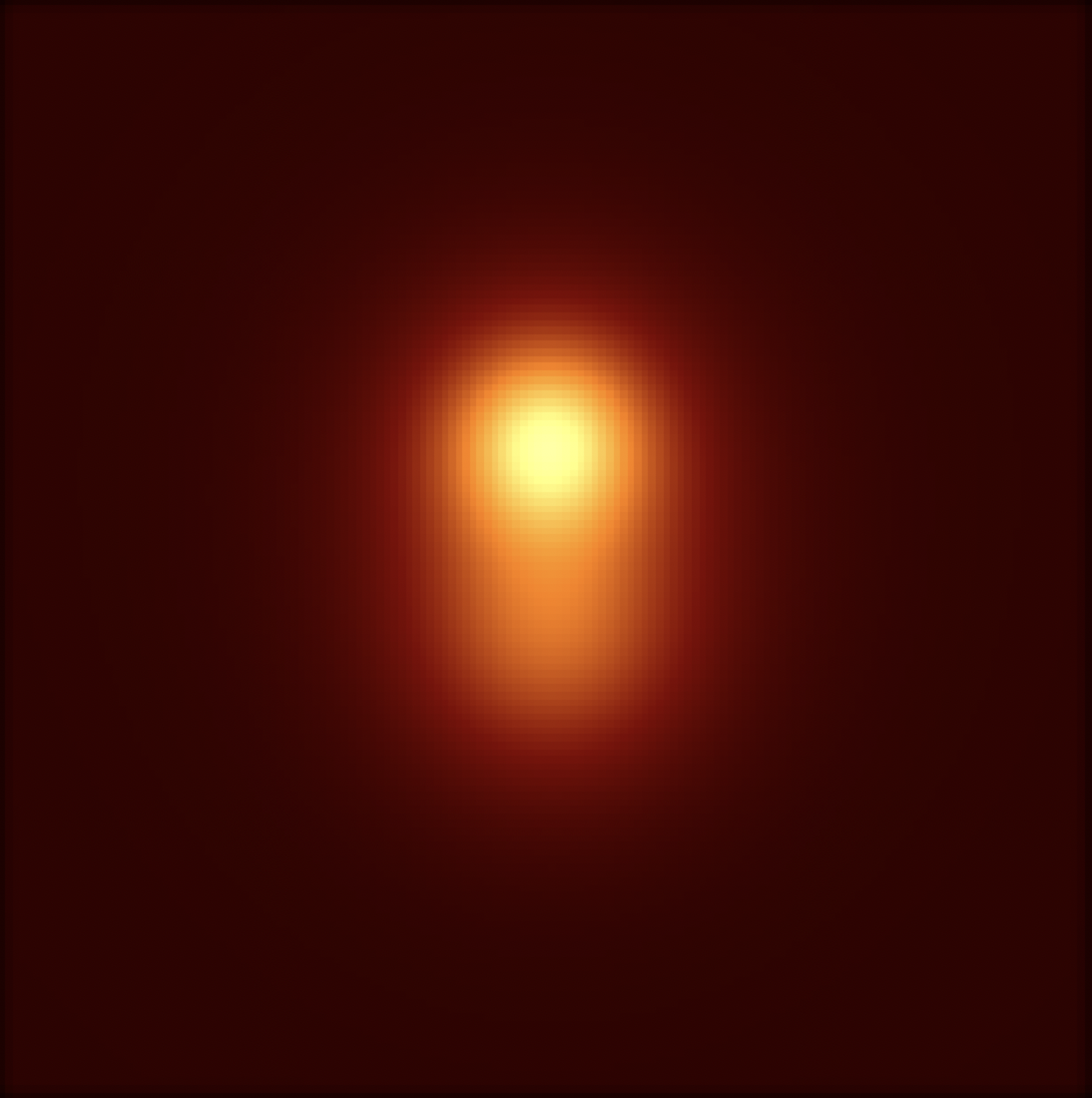}
           \includegraphics[scale=0.3]{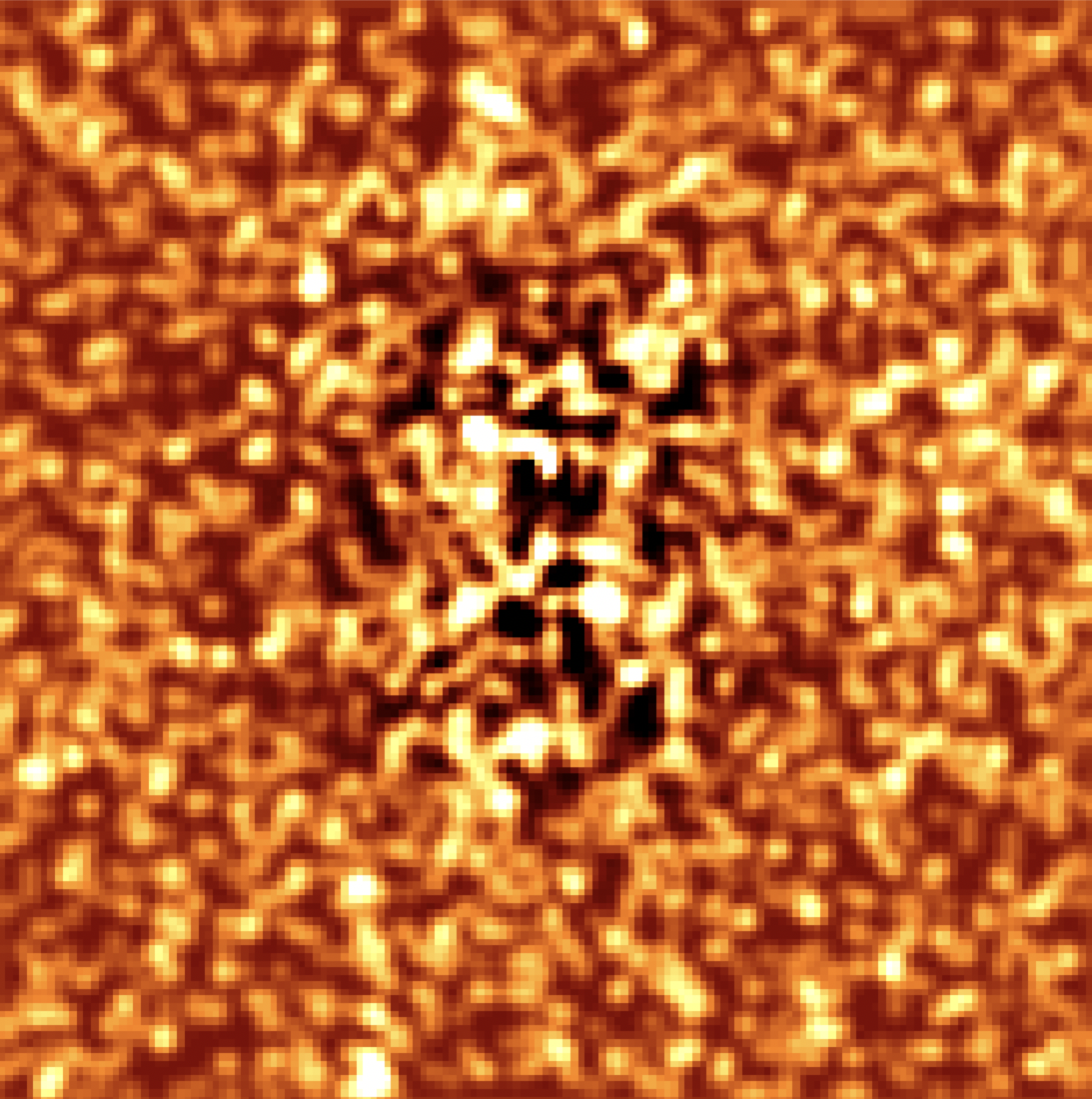}}
            \caption{\emph{Left Panel}: AGN-subtracted \emph{Chandra} X-ray image of \texttt{XDCP0044}. The magenta circles represent the north clump and south clump identified in \cite{Tozzi2014} with two different temperatures} and the cyan crosses represent the four complexes identified in this cluster, \texttt{Complex A} and \texttt{Complex B} \citep{Travascio2020} in the north and \texttt{Complex M} and \texttt{Complex N} in the south. The green crosses represent the X-ray emission peaks found with Sherpa image fitting. \emph{Central Panel}: Best-fit model obtained with \emph{Sherpa} of the AGN-subtracted X-ray image. \emph{Right Panel}: Residuals obtained with \emph{Sherpa}.
         \label{fig:sherpa}
\end{figure*}

From the analysis of the suface brightness distribution of \texttt{XDCP0044}, \cite{Tozzi2014} found that this cluster shows the presence of two clumps (see Figure \ref{fig:sherpa} \emph{left panel}, magenta circles) with possibly different temperatures, the north clump with $kT=7.1^{+1.3}_{-1.0}$ keV and the south clump with $kT=5.5^{+1.2}_{-1.0}$ keV, which are consistent at 
slightly less than $2\sigma$. To confirm the presence of two different clumps, we analyze the \emph{Chandra} X-ray images with the \emph{Sherpa} software in {\tt ciao}. 

We consider the AGN-subtracted \emph{Chandra} X-ray image in the soft band and fit the data in this region with a double two-dimensional beta model plus background using Cash statistic, in order to find the emission peak of each clump. 
We find that the first peak is associated to the cluster core studied by \cite{Travascio2020} (north clump), while the second peak is located south the cluster core, shifted with respect to the region studied in this work, as shown in Figure \ref{fig:sherpa} (green and cyan cross, respectively, in the \emph{left panel}). To better understand the results obtained from the analysis, we show the best-fit model and the residuals obtained with \emph{Sherpa} in Figure \ref{fig:sherpa} (\emph{central panel} and \emph{right panel}, respectively). 
We inspect the residuals and the SNR ratio map of the residuals to search for possible 
structure in the X-ray emission not accounted for by our double beta model.  We do not find any
significant residuals, expect for a weak hint of a slightly brighter arc in the overlapping region between the two clumps.  This shows 
that the two halo model is more than adequate to describe the
data, and suggesting that further signs of the ongoing merger should be investigated with 
future, high resolution X-ray missions such as Lynx \citep{2018LynxTeam} 
and AXIS \citep{2019Mushotzky,2020Marchesi}.

To summarize, we note that the main halo \citep[hosting the core studied by][]{Travascio2020} is significantly brighter and can be considered 
virialized with an almost formed BCG at the center. Moreover, we note that the southern clump 
has a lower surface brightness, and that its X-ray centroid is far from \texttt{Complex M} and \texttt{Complex N}.  However, the position of the X-ray centroid of the 
weaker halo may have been biased by the low signal.  Finally, 
we note another feature consisting in a weak, but
clearly visible extension toward the west, suggesting a very complex dynamical state
that has not been modeled in the current analysis.

Despite our data set is not sufficient to provide a comprehensive
explanation of the dynamical state of \texttt{XDCP0044}, we argue that
we may be observing the merger of two
halos, both recently virialized
or in the process of being virialized, and each one with its own BCG in the 
assembling phase.  We do expect that the two systems 
identified in this work, \texttt{Complex M} and \texttt{Complex N}, 
will merge within $\sim$ 370 Myr to form a massive galaxy ($M_{*}\sim10^{11}M_{\odot}$). Following the same procedure applied in Section \ref{Mass assembly}, we suggest that these complexes will eventually join the massive galaxy formed by \texttt{Complex A} and \texttt{Complex B} in the core \citep{Travascio2020} on a timescale of $\sim$ 6 Gyr, which is in agreement with simulations \citep{BoylanKolchin2008, Villalobos2013}.

\section{Summary $\&$ Conclusions}
\label{sec:conclusioni}

Dense regions in high-z galaxy clusters are ideal laboratories to investigate the interplay between galaxies, nuclear activity and the intergalactic gas. In particular, we have analyzed high resolution multi-band HST images, NIR IFU KMOS spectroscopic data and the X-ray \emph{Chandra} data of a small (24 kpc $\times$ 24 kpc) but very dense region around one of the X-ray point-like sources discovered within the galaxy cluster \texttt{XDCP0044} at z$\sim$1.6, located $\sim$ 157 kpc from its center. 
The main results of this paper can be summarized as follows:

\begin{enumerate}
\item thanks to the high resolution HST data, we found that the analyzed region is denser than expected from the analysis of ground based observations. In particular, up to 9 sources have been detected in only 24 kpc $\times$ 24 kpc. This is even denser than what found in the core of the cluster by \cite{Travascio2020} and in the core of the Spiderweb protocluster at $z\sim 2.156$. We have performed the photometric analysis of all the available HST bands, i.e. F105W, F140W and F160W, and computed the fluxes and magnitudes of the 9 identified sources. These values have been then used to compute the monochromatic luminosities at $5100\text{\AA}$.
\item From the KMOS IFU spectroscopic analysis, we found that at least 6 of the 9 sources detected in HST show a narrow H$\alpha$ emission line, and two of them also the H$\beta$ emission line. From the derived redshifts, which range from 1.5728 to 1.5762, we find that all 6 sources are confirmed cluster members.
Moreover, the sources seems to be located in two different subgroups: \texttt{Complex M} and \texttt{Complex N} at a projected distance d$\sim$13 kpc. \texttt{Complex M} is formed by sources \texttt{m1}, \texttt{m2} and \texttt{m3}, with redshift ranging from 1.5756 to 1.5762, and \texttt{Complex N} by sources \texttt{n1}, \texttt{n2} and \texttt{n3}, with redshift ranging from 1.5728 to 1.5732.
\item IFU spectroscopic data allow us to study the kinematic of the analyzed complexes, through velocity shift maps of the narrow H$\alpha$ emission line. From such maps, we found that \texttt{Complex N} has a negative velocity shift ($\sim$ $300$ km/s) with respect to \texttt{Complex M}. Moreover, from the velocity shift maps, we found significant emission below the source \texttt{m1}, at the same wavelength of the narrow H$\alpha$ emission line of \texttt{m1}, that can be interpreted either as a diffuse gas in the midst of galaxies, i.e. gas stripped during the merger, or as an ENLR around source \texttt{m1}. We also find hints of emission between the sources \texttt{m1} and \texttt{n3}, with a high velocity shift ($\sim$ $450$ km/s).
\item From the \emph{Chandra} X-ray spectrum of source \texttt{m1}, we find a photon index between 1.8 and 2.0, typical of an AGN, and an obscuration of $N_{H} \sim 10^{22}cm^{-2}$ for an intrinsic X-ray luminosity of $L_{[2-10]keV} \sim 10^{43}$ erg/s.
We detected a broad H$\alpha$ emission line in the spectrum of source \texttt{m1}. This source is therefore classified as a BLAGN. By applying the virial formula, using the measured H$\alpha$ $FWHM>1500$ km/s and luminosity, we estimate a SMBH mass of $M_{BH}\sim 10^{7}M_{\odot}$.
\item The 5100$\text{\AA}$ luminosity of the 6 confirmed cluster members have been derived by interpolating the computed HST fluxes at F105W and F140W. Source \texttt{m1}, with $\rm Log[L_{5100\text{\AA}}/erg/s] \sim 43.85$, is the most powerful source, in agreement with the fact that it hosts an AGN. Moreover, by applying to \texttt{m1} the AGN bolometric corrections by \cite{Krawczyk2013}, we estimated the bolometric luminosity using $L_{5100\text{\AA}}$, finding $\rm Log[L_{bol}/erg/s] \sim 44.49$. This value is in agreement with what found by applying the bolometric corrections by \cite{Duras2020} to the X-ray $[2-10]$keV luminosity, confirming that the monochromatic luminosity at $5100\text{\AA}$ is related to a nuclear component.
\item Finally, from the BH mass and the bolometric luminosity, we derived the Eddington luminosity, $\rm Log[L_{Edd}/erg/s] \sim 45.2$, and a relatively high Eddington ratio  $\lambda_{Edd} \sim 0.2$.

\item While the merging time scale is estimated to be 10 Myr for 
\texttt{Complex M} and 30 Myr for \texttt{Complex N}, the two complexes, at a distance of 
$\sim$13 kpc with a velocity shift of $\sim$300 km/s, 
will likely merge within a timescale comparable to dynamical friction 
to form a massive galaxy, possibly the second BCG of \texttt{XDCP0044}.

\item After a revised analysis of the X-ray image of \texttt{XDCP0044}, we conclude that
the two complex pairs (\texttt{A}+\texttt{B}, \cite{Travascio2020}, and \texttt{M}+\texttt{N}) are most likely associated to two 
merging halos, with different properties in term of average ICM temperature and surface brightness.

\end{enumerate}

To summarize, we investigated a 
region in the high-redshift galaxy cluster 
\texttt{XDCP0044}, suggesting that it is probably the formation site  of one of the two BCGs of the cluster that is undergoing a major merger. These results corroborate a scenario in which AGN activity is triggered during the mergers between gas-rich galaxies which provide the fuel for both the AGN and the starburst activity. Given the high BH accretion rate and the high SF rate, the BH is expected to rapidly increase its mass during the galaxies’ mergers and later, through feedback, quench the SF, leading to the formation of a massive and passive galaxy, as observed in local galaxy clusters. Deeper IR integral field spectroscopy may provide further insights into the dynamical
state of the galaxy systems, and help in constraining their future evolution. As for the hot, diffuse baryons, the loss of sensitivity of the only arcsec resolution X-ray facility ({\sl Chandra}) makes it impossible to go deeper in the X-ray band.  However, high-resolution imaging of the SZ effect due to the ICM itself may be obtained thanks to ALMA. Also, thanks to the advent of JWST, detailed studies of the assembling galaxies in highz clusters will become more frequent, allowing us to reach a comprehensive description of a crucial phase in the formation and evolution of massive galaxies.

\begin{acknowledgements}
    We thank G. De Lucia and E. Rasia for useful discussions, and the anonymous referee for the accurate comments on the paper.
\end{acknowledgements}

\bibliographystyle{aa}
\bibliography{paper}



%
%
\end{document}